\documentclass[aps,prl,twocolumn,superscriptaddress,longbibliography,floatfix,10pt]{revtex4-2}

\usepackage{float,amsmath,amssymb,bbm,mathrsfs,bm,braket,color,graphicx,comment,amsfonts,dsfont,mathrsfs}
\usepackage{xcolor}
\usepackage[colorlinks,citecolor=blue,urlcolor=blue]{hyperref}
\usepackage[mathcal]{euscript}
\usepackage[normalem]{ulem}
\usepackage{comment}
\usepackage{dsfont}
\usepackage{xfrac}
\usepackage{graphicx} 

\DeclareMathAlphabet\mathbfcal{OMS}{cmsy}{b}{n}

\begin{document}
\title{Controlling photo-excited electron-spin\\
 by light-polarization in ultrafast-pumped altermagnets}
\author{Amir Eskandari-asl}
\affiliation{Dipartimento di Fisica E.R. Caianielloâ Universita degli Studi di
Salerno, I-84084 Fisciano (SA), Italy}
\author{Jorge I. Facio}
\affiliation{Centro Atomico Bariloche and Instituto Balseiro, CNEA, 8400 Bariloche,
Argentina}
\affiliation{Instituto de Nanociencia y Nanotecnologia CNEA-CONICET, 8400 Bariloche
Argentina}
\author{Oleg Janson}
\affiliation{Leibniz Institute for Solid State and Materials Research, IFW Dresden,
Helmholtzstra\ss e 20, 01069 Dresden, Germany}
\author{Adolfo Avella}
\email{a.avella@unisa.it}

\affiliation{Dipartimento di Fisica E.R. Caianiello, Universita degli Studi di
Salerno, I-84084 Fisciano (SA), Italy}
\affiliation{CNR-SPIN, Unit di Salerno, I-84084 Fisciano (SA), Italy}
\affiliation{CNISM, Unit di Salerno, Universita degli Studi di Salerno, I-84084
Fisciano (SA), Italy}
\author{Jeroen van den Brink}
\email{j.van.den.brink@ifw-dresden.de}

\affiliation{Leibniz Institute for Solid State and Materials Research, IFW Dresden,
Helmholtzstra\ss e 20, 01069 Dresden, Germany}
\affiliation{W\"urzburg-Dresden Cluster of Excellence ct.qmat, Technische Universit\"at
Dresden, 01062 Dresden, Germany}
\date{\today}
\begin{abstract}
Altermagnets (AMs) constitute a novel class of spin-compensated materials
in which the symmetry connecting opposite-spin sublattices involves
a spatial rotation. Here, we uncover a set of unique non-linear, light-driven
properties that set AMs apart from traditional ferro- and antiferromagnets.
We demonstrate theoretically that the \textit{polarization} of an
electromagnetic pulse that photo-excites electrons and holes in an
AM, controls the \textit{spin orientation} of these non-equilibrium
charge carriers. For a $d$-wave AM model and a prototype material, we
show that very large post-pump spin polarizations may be attained
by exploiting resonances. We show that this protocol also allows,
in an AM, to directly probe the spin splitting of the electronic states
in energy and momentum space. Thus, it can be used to identify and
characterize altermagnetic materials via ultrafast pump-probe Kerr/Faraday
spectroscopy or spin- and time-resolved ARPES. This opens up the possibility
of devising ultrafast optical switches of non-equilibrium spin-polarization,
finely tunable by adjusting the pump-pulse characteristics. 
\end{abstract}
\maketitle
Altermagnetism has recently emerged as a new type of magnetic ordering,
distinct from ferro- and antiferromagnetism. Similarly to antiferromagnets~(AFMs),
the net magnetization of altermagnets~(AMs) vanishes by symmetry.
They differ from AFMs because the enforcing symmetry that connects
the two magnetic sub-lattices is not merely an inversion or translation,
but also involves a rotation~\cite{Smejkal20,Smejkal22}. This leads
to a breaking of the spin degeneracy of the electronic states in their
non-relativistic band structure with an energy scale set by the local
exchange field, which is generally much larger than the relativistic
spin-orbit-coupling energy scale.  A large number of AMs have already been identified~\cite{Smejkal22a,Naka19,Yuan20,Naka21,Guo23}, and this number continues to grow.
Various linear response properties of AMs have been recognized that
may render them of practical interest, for instance, as
spin current generators relevant for spintronics~\cite{Smejkal20,Smejkal22}.
Also, the recently developed Landau theory of altermagnetism~\cite{McClarty24}
allows to relate the formation of altermagnetic order directly to
other key linear response properties such as anomalous Hall conductivity~\cite{Smejkal20,Sato24},
Edelstein response~\cite{Trama24,Hu24} and piezomagnetic~\cite{Consoli21,Aoyama24,Yershov24},
magneto-optic, and magneto-elastic~\cite{Yershov24_2} effects.
In particular, the linear magneto-optical response in the presence of an external magnetic field, which involves momentum- and spin-dependent matrix elements, is sensitive to the presence of altermagnetism and depends on the orientation of the ground state N{\'e}el vector~\cite{Vila24}.

Here, we take a step beyond the linear responses of AMs and consider their
non-linear, light-driven properties. In general, the possibility of
controlling the physical properties of solids with ultrafast electromagnetic
(EM) pulses is a fascinating goal at the core of a broad research
field~\cite{brabec2000intense,schmitt2008transient,krausz2009attosecond,krausz2014attosecond,Zurch_17,borrego2022attosecond,inzani2022field}.
In this Letter, we set out to determine how the features of a strong
EM pulse (its polarization, frequency, amplitude, and duration) affect
the post-pump photo-induced charge carriers (their density, spin and
momentum) in AMs. Using the Dynamical Projective Operatorial Approach
(DPOA)~\citep{inzani2022field,eskandari2024time,eskandari2024generalized},
which allows to efficiently study out-of-equilibrium systems by projecting
the time-dependent operators on their equilibrium counterparts, we
analyze the field-induced charge carriers in a $d$-wave AM model
as well as a metallic RuO$_{2}$ bilayer, employing its electronic
properties we calculated \textit{ab initio}.

Our analysis reveals that by changing the linear polarization of the
EM pump pulse, it is possible to photo-excite electrons with a specific
spin orientation, exploiting the resonance of specific momentum regions
of the band dispersion to the pulse frequency and, in particular,
the polarization and momentum dependence of the light-matter coupling
per spin direction in AMs. We show how the effect can be tuned and
optimized adapting the pulse characteristics and demonstrate how the
large electronic spin-splittings of AMs in momentum space can be leveraged
to generate large densities of pump-induced spin polarization. The
pulse polarization control of the spin direction and net magnetization
of photo-pumped charge carriers are underlain by the altermagnetic
symmetry and thus absent in conventional ferro- or antiferromagnets.

\begin{figure}
\centering{\includegraphics[width=8cm]{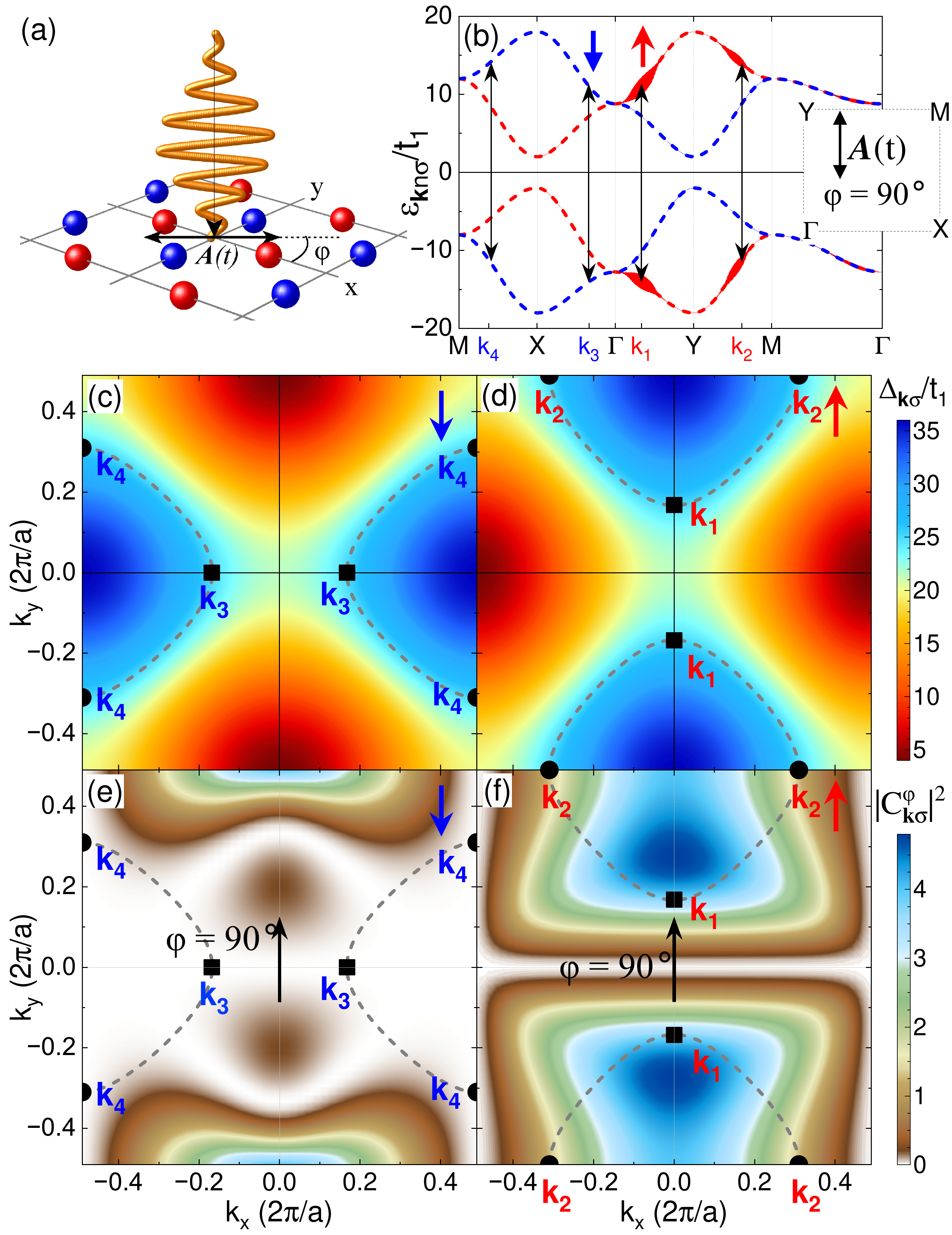}} \caption{(a) 
Sketch of the $d$-wave AM model and the vector potential $\boldsymbol{A}\left(t\right)$
of the pump pulse impinging on it. (b) Spin-up (red) and spin-down
(blue) electronic bands. The thickness of the superimposed solid lines
is proportional to the post-pump electron and hole photo-excited populations,
for $A_{0}=0.2\frac{h}{ea}$, $\omega_{\mathrm{pu}}=25\frac{t_{1}}{\hbar}$,
$\tau_{\mathrm{pu}}=0.8\frac{\hbar}{t_{1}}$, and $\varphi=90{^{\circ}}$.
Momenta $k_{i}$, for $i=1,\ldots,4$, denote \emph{resonant} $\boldsymbol{k}$-points.
Only $k_{1,2}$ host non-negligible post-pump photo-excited populations
because of the corresponding noticeable coupling to the pump pulse.
(c-d) Band gap $\Delta_{\boldsymbol{k}\sigma}$ over the whole BZ
for spin-down and spin-up bands, respectively. The dashed lines mark
the positions of \emph{resonant} $\boldsymbol{k}$-points. (e-f) Light-matter
coupling strengths, $\left|C_{\boldsymbol{k}\sigma}^{\varphi}\right|^{2}$,
over the whole BZ for spin-down and spin-up bands, respectively,
and $\varphi=90{^{\circ}}$.}
\label{fig:1_system} 
\end{figure}

\textit{$d$-wave AM model Hamiltonian}---To capture the photo-excitation
behavior of an AM induced by a linearly polarized EM pulse, we first
consider the simple realization of a $d$-wave AM introduced in Ref.~\cite{antonenko2024}.
The model takes the form of a 2D Lieb lattice shown in Fig.~\ref{fig:1_system}(a),
where the anti-parallel magnetic moments (blue and red dots) form
two distinct sublattices related by fourfold rotation. The AM character
emerges from the two sublattices having distinct local environments
due to the presence of further non-magnetic sites~\cite{antonenko2024}.
A Kondo-type interaction couples frozen localized spins with itinerant
electrons hopping only among the two magnetic sublattices, resulting
in a spin-split band structure, see Fig.~\ref{fig:1_system}(b).
For simplicity, we take a parametrization (see the End Matter~A\ref{sec:AM-model})
that comes with a clean gap between the AM bands at half filling,
see Fig.~\ref{fig:1_system}(b). As expected, the spin-up, $\varepsilon_{\boldsymbol{k}n\uparrow}$,
and spin-down, $\varepsilon_{\boldsymbol{k}n\downarrow}$, bands are
related by a $\pi/2$ rotation in the $\boldsymbol{k}$ space {[}see Fig.~\ref{fig:1_system}
(b){]}. We indicate the valence (conduction) band by $n=1$(2). 
The spin-specific band gap, $\Delta_{\boldsymbol{k}\sigma}=\varepsilon_{\boldsymbol{k}2\sigma}-\varepsilon_{\boldsymbol{k}1\sigma}$,
naturally exhibits the AM symmetry {[}see Figs.~\ref{fig:1_system}
(c-d){]}. 

\textit{Light pulse impinging on a solid}---To determine how an
ultrafast EM pulse impacts the spin and charge distribution of photo-excited
states in an AM, we start from the spin-conserving dipole-gauge Hamiltonian
describing the interaction between a classical EM pulse in the dipole
approximation and a solid-state lattice system~\citep{schuler2021gauge,eskandari2024time,eskandari2024generalized}
\begin{multline}
\hat{H}=\sum_{\boldsymbol{k},\nu,\nu^{\prime},\sigma}c_{\boldsymbol{k}\nu\sigma}^{\dagger}T_{\boldsymbol{k}+\frac{e}{\hbar}\boldsymbol{A}\left(t\right),\nu\nu^{\prime}\sigma}c_{\boldsymbol{k}\nu^{\prime}\sigma}\\
+e\boldsymbol{E}\left(t\right)\cdot\sum_{\boldsymbol{k},\nu,\nu^{\prime},\sigma}c_{\boldsymbol{k}\nu\sigma}^{\dagger}\boldsymbol{D}_{\boldsymbol{k}+\frac{e}{\hbar}\boldsymbol{A}\left(t\right),\nu\nu^{\prime}\sigma}c_{\boldsymbol{k}\nu^{\prime}\sigma},\label{eq:H-A-DG}
\end{multline}
where $\boldsymbol{k}$ is the crystal momentum that is summed over
the Brillouin zone (BZ), $\sigma$ is the electronic spin,
$\nu$ and $\nu^{\prime}$ are sets of quantum numbers identifying
orthogonal localized electronic states (e.g., the maximally localized
Wannier states), $c_{\boldsymbol{k}\nu\sigma}$ ($c_{\boldsymbol{k}\nu\sigma}^{\dagger}$)
is the annihilation (creation) operator of an electron in the state
with quantum numbers $\left(\boldsymbol{k},\nu,\sigma\right),$ $\boldsymbol{A}\left(t\right)$
is the vector potential and $\boldsymbol{E}\left(t\right)=-\partial_{t}\boldsymbol{A}\left(t\right)$
is the electric field of the impinging pulse. $T_{\boldsymbol{k}\nu\nu^{\prime}\sigma}$
is the matrix element of the first-quantization (single-particle)
equilibrium Hamiltonian and $\boldsymbol{D}_{\boldsymbol{k}\nu\nu^{\prime}\sigma}$
is the matrix element of the local dipole in the reciprocal space.

\begin{figure}[t]
\centering{\includegraphics[width=8cm]{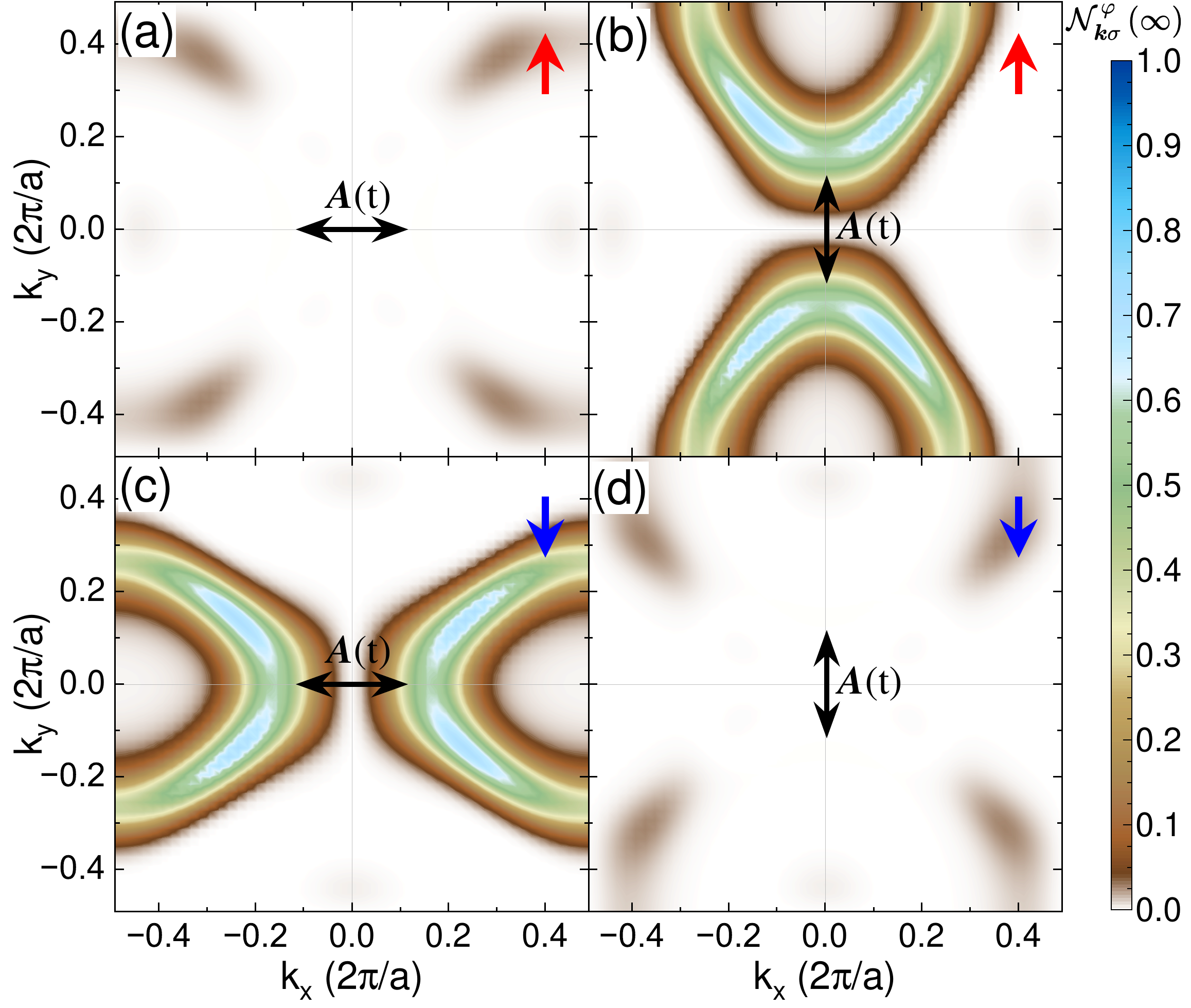}}
\caption{The pump pulse polarization controls the spin of the photo-excited
charge carriers: shown are the post-pump spin up/down electron populations,
$\mathcal{N}_{\boldsymbol{k}\sigma}^{\varphi}\left(\infty\right)$,
for pump pulse polarization $\varphi=0/90{^{\circ}}$: (a,b) spin-up
, (c,d) spin-down, (a,c) $\varphi=0{^{\circ}}$, and (b,d) $\varphi=90{^{\circ}}$.}
\label{fig:2_N_toymodel} 
\end{figure}

\textit{Mechanism for polarization control of post-pump {electron}-spin 
distribution}---Before presenting the calculated post-pump, non-equilibrium 
{electron}-spin distribution, we first elucidate the mechanism by which 
the polarization of the {light} pulse can control and switch the spin orientation 
of the excited electrons, a property that sets AMs apart from conventional FMs and AFMs. 
We consider a pump-pulse with a polarization in the lattice plane parametrized
by the angle $\varphi$ {formed} with the x axis, as indicated in Fig.~\ref{fig:1_system} (a).
The cross-section for photo-excitation is large when the spin-specific band gap 
$\Delta_{\boldsymbol{k}\sigma}$ is in resonance with the pump-pulse frequency 
$\omega_{\mathrm{pu}}$. The resonant momenta are indicated by dashed lines 
in Fig.~\ref{fig:1_system}(c-f) and include, for instance, momenta 
$\boldsymbol{k}_{1-4}$ indicated in Fig.~\ref{fig:1_system}(b). 
Here $\boldsymbol{k}_{1,2}$ correspond to resonances of spin-up electrons, 
and $\boldsymbol{k}_{3,4}$ to spin-down ones. It is important to note that 
due to the altermagnetic symmetry, $\boldsymbol{k}_{1,2}$ are related to 
$\boldsymbol{k}_{3,4}$ by a $\pi/2$ rotation. The light-matter coupling 
strength $C_{\boldsymbol{k}\sigma}^{\varphi}$ determines how strongly 
these states couple to the {light} field.
In an AM, this essential ingredient depends in a non-trivial manner 
on the spin $\sigma$ of the electronic states involved and the 
polarization $\varphi$ of the {light pulse}. In the following, 
we will determine {the expression of} $C_{\boldsymbol{k}\sigma}^{\varphi}$ 
and its dependence on {light}-pulse characteristics explicitly. 
The calculated light-matter coupling strength for the d-wave AM model 
is shown in Fig.~\ref{fig:1_system}(e,f) for $\varphi=90{^{\circ}}$. 
For this polarization direction, one observes that {although} 
$\boldsymbol{k}_{3,4}$ are in resonance, these states only couple weakly 
to the {light} field so that 
the post-pump density of spin-down states {is} 
low (see Fig.~\ref{fig:2_N_toymodel}(d)). However, {for} 
the same polarization {direction}, $\boldsymbol{k}_{1,2}$ 
strongly couple {to the light field} (Fig.~\ref{fig:1_system}(f)) and 
{host} a large spin-up density after pumping (Fig.~\ref{fig:2_N_toymodel}(b)). 
Because of the altermagnetic symmetry, the rotation of the incoming 
{light-pulse} polarization by $\pi/2$ exchanges the role of 
spin up and spin down, see Fig.~\ref{fig:2_N_toymodel}, 
giving rise to spin-switching. 

\textit{Dynamical projective operatorial approach (DPOA)}---To determine 
quantitatively the impact of the pump pulse on the electronic state of
the system, we use the DPOA \citep{inzani2022field,eskandari2024time,eskandari2024generalized},
which efficiently and effectively study out-of-equilibrium systems
by projecting the time-dependent operators on their equilibrium counterparts,
$c_{\boldsymbol{k}n\sigma}\left(t\right)=\sum_{n^{\prime}}P_{\boldsymbol{k}nn{}^{\prime}\sigma}\left(t\right)c_{\boldsymbol{k}n^{\prime}\sigma}$,
and moving the solution of the operatorial dynamics to the solution
of the equations of motion of the dynamical projection matrices, 
$P_{\boldsymbol{k}nn{}^{\prime}\sigma}\left(t\right)$
(see the End Matter~B\ref{sec:End-material-DPOA}). In principle,
this allows to compute any out-of-equilibrium response, such as time-resolved
ARPES \citep{eskandari2024time} and transient optical properties
\citep{eskandari2024generalized}. In particular, the electronic band
populations can be obtained as $N_{\boldsymbol{k}n\sigma}\left(t\right)=\left\langle c_{\boldsymbol{k}n\sigma}^{\dagger}\left(t\right)c_{\boldsymbol{k}n\sigma}\left(t\right)\right\rangle =\sum_{n^{\prime}}P_{\boldsymbol{k}nn{}^{\prime}\sigma}\left(t\right)f_{+}\left(\varepsilon_{\boldsymbol{k}n^{\prime}\sigma}\right)P_{\boldsymbol{k}nn{}^{\prime}\sigma}^{\star}\left(t\right)$,
where $f_{+}\left(\varepsilon\right)$ is the Fermi distribution function
\citep{eskandari2024time} (see the End Matter~B\ref{sec:End-material-DPOA}).

\textit{Resulting post-pump {electron-}spin distribution}---The pump pulse 
that we consider features a vector potential $\boldsymbol{A}\left(t\right)=A_{0}e^{-\left(4\ln2\right)t^{2}/\tau_{\mathrm{pu}}^{2}}\sin\left(\omega_{\mathrm{pu}}t\right)\hat{\boldsymbol{A}}$, where $A_{0}$ is its amplitude, $\tau_{\mathrm{pu}}$ is the FWHM of its Gaussian envelope centered at time $t=0$, and $\omega_{\mathrm{pu}}$ is its central frequency. 
%
The electronic excited population, which is the positive excess post-pump
electronic population with respect to the thermal equilibrium one
summed over all bands, is denoted by $\mathcal{N}_{\boldsymbol{k}\sigma}^{\varphi}\left(\infty\right)$.
Given
that the main excitation processes in the model are due to one-photon
resonances, the dimensionless light-matter coupling is determined
from the off-diagonal element of the \emph{velocity} (first-order)
term in the Peierls expansion: $C_{\boldsymbol{k}\sigma}^{\varphi}=\frac{1}{t_{1}a}\sum_{\nu\nu^{\prime}}\Omega_{\boldsymbol{k}1\nu^{\prime}\sigma}^{\dagger}\left(\partial_{k_{A}}T_{\mathbf{k}\nu^{\prime}\nu\sigma}\right)\Omega_{\boldsymbol{k}\nu2\sigma}$
(see End Matter~B\ref{sec:End-material-DPOA}). 
$\left|C_{\boldsymbol{k}\sigma}^{\varphi}\right|^{2}$ exhibits the
same symmetry of $\varepsilon_{\boldsymbol{k}n\sigma}$ and $\Delta_{\boldsymbol{k}\sigma}$
under rotation of the polarization. As indicated above, for fixed polarization,
$\left|C_{\boldsymbol{k}\sigma}^{\varphi}\right|^{2}$ is instead
very different between spin up and spin down {[}see Figs.~\ref{fig:1_system}
(e-f){]}. 

The post-pump electron photo-excited populations, $\mathcal{N}_{\boldsymbol{k}\sigma}^{\varphi}\left(\infty\right)$,
of the $d$-wave AM model for the four relevant cases obtained by
crossing the values of spin $\sigma$ (up and down) and polarization
$\varphi$ ($0{^{\circ}}$ and $90{^{\circ}}$) are shown in Fig.~\ref{fig:2_N_toymodel}.
The noticeable difference
in the coupling strength $\left|C_{\boldsymbol{k}\sigma}^{\varphi}\right|^{2}$
between spin up and spin down for the chosen pump pulse frequency
and polarization ($\varphi=90{^{\circ}}$), that is along the dashed
lines of Figs.~\ref{fig:1_system} (e) and (f), leads to the huge
difference between the values of $\mathcal{N}_{\boldsymbol{k}\uparrow}^{\varphi}\left(\infty\right)$
and $\mathcal{N}_{\boldsymbol{k}\downarrow}^{\varphi}\left(\infty\right)$
in Figs.~\ref{fig:2_N_toymodel} (b) and (d), and determines the
regions in $\boldsymbol{k}$ space where one can find the higher values
of $\mathcal{N}_{\boldsymbol{k}\uparrow}^{\varphi}\left(\infty\right)$:
the momentum and spin selectivity by polarization is demonstrated.
The symmetry of $\mathcal{N}_{\boldsymbol{k}\sigma}^{\varphi}\left(\infty\right)$
relates Figs.~\ref{fig:2_N_toymodel} (a) and (c), to (b) and (d). 

\begin{figure}[t]
\centering{\includegraphics[width=1\columnwidth]{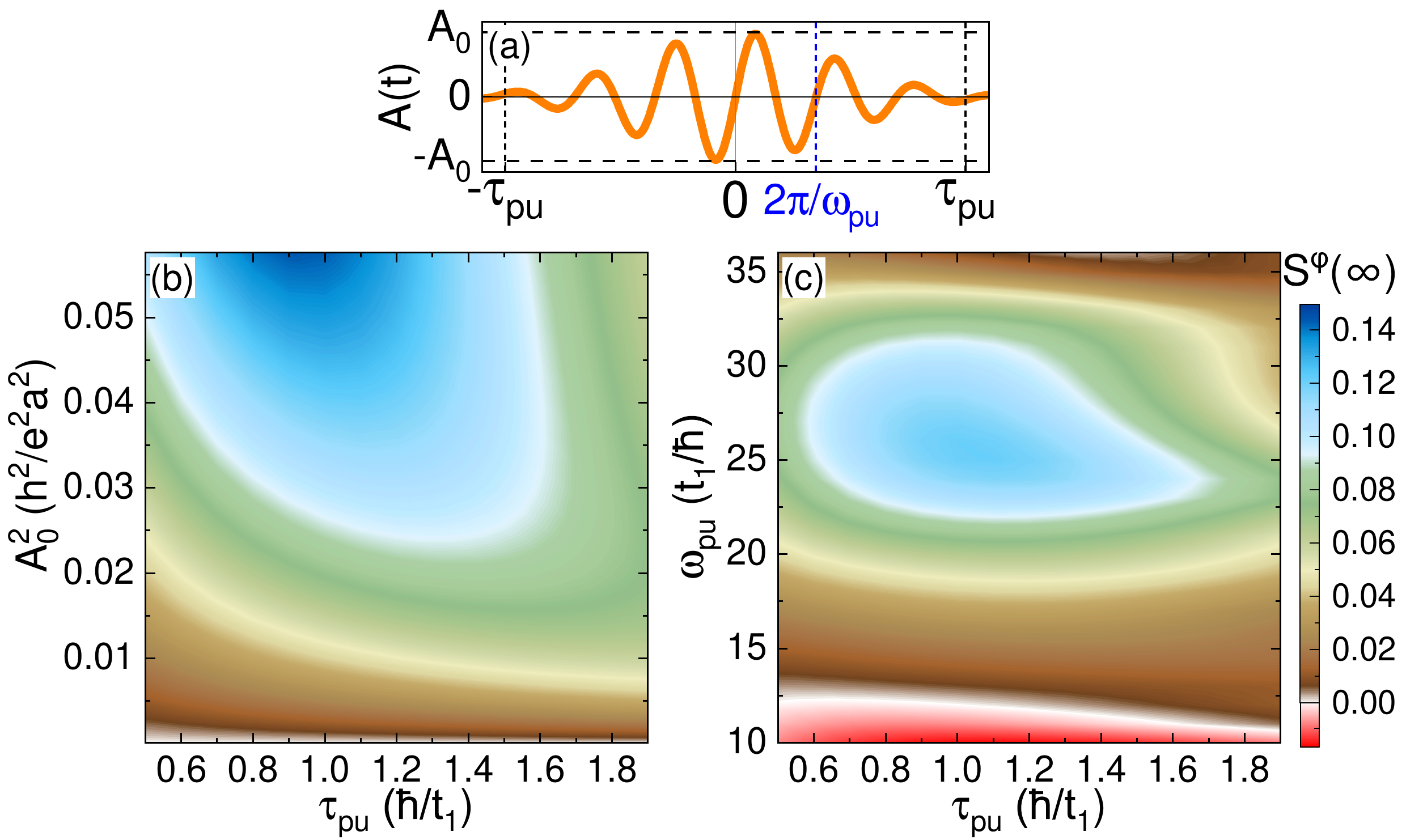}}\caption{(a) 
Characteristics of the vector potential of the pump pulse. (b-c)
Post-pump electron photo-excited magnetization per unit cell, $S^{\varphi}\left(\infty\right)$,
as a function of the FWHM, $\tau_{\mathrm{pu}}$, and (b) the square
of the pump pulse amplitude, $A_{0}^{2}$, (c) the pump pulse frequency,
$\omega_{\mathrm{pu}}$, for $\varphi=90{^{\circ}}$ and the same
parameter values of Fig.~\ref{fig:1_system}. }
\label{fig:3_N_Aw} 
\end{figure}

Having established that the pump pulse frequency, $\omega_{pu}$, 
selects the \emph{resonant} momentum region, and the pump pulse 
polarization, $\varphi$, is very effective in selecting the spin, 
we turn to the effects of the pulse's amplitude, $A_{0}$, and duration, measured
through the FWHM, $\tau_{\mathrm{pu}}$, of its Gaussian envelope
{[}see Fig.~\ref{fig:3_N_Aw} (a){]}.
In Figs.~\ref{fig:3_N_Aw} (b) and (c), we report the post-pump electron
photo-excited magnetization per unit cell, $S^{\varphi}\left(\infty\right)=\frac{a^{2}}{4\pi^{2}}\int_{BZ}\left[\mathcal{N}_{\boldsymbol{k}\uparrow}^{\varphi}\left(\infty\right)-\mathcal{N}_{\boldsymbol{k}\downarrow}^{\varphi}\left(\infty\right)\right]d\boldsymbol{k}$,
to discuss its dependence on pump pulse FWHM, amplitude, and frequency
at fixed polarization ($\varphi=90{^{\circ}}$).

Within the overall time span of the pump pulse, the electron photo-excited
populations at (and close to) \emph{resonant} $\boldsymbol{k}$-points
undergo Rabi-like oscillations with a Rabi frequency, $\omega_{R,\boldsymbol{k}\sigma}^{\varphi}$,
which is proportional to $\left|C_{\boldsymbol{k}\sigma}^{\varphi}\right|A_{0}$
\citep{eskandari2024dynamical,inzani2023field}. Accordingly, the
post-pump electron photo-excited populations, $\mathcal{N}_{\boldsymbol{k}\sigma}^{\varphi}\left(\infty\right)$,
is roughly proportional to 
$\sin^{2}(\omega_{R,\boldsymbol{k}\sigma}^{\varphi}\tau_{\mathrm{pu}}/2)$,
which can be approximated to $(\omega_{R,\boldsymbol{k}\sigma}^{\varphi}\tau_{\mathrm{pu}}/2)^2$
%
%
for small enough pump pulse amplitudes and durations \citep{eskandari2024dynamical}.
This explains why, in Fig.~\ref{fig:3_N_Aw} (b), $S^{\varphi}\left(\infty\right)$
increases monotonically with the pump pulse FWHM, $\tau_{\mathrm{pu}}$,
only at small enough amplitudes, $A_{0}$, while it has a re-entrant
behavior on continuously increasing $\tau_{\mathrm{pu}}$ for large
enough values of $A_{0}$.

On varying the pump pulse frequency, $\omega_{pu}$, instead, we change
the loci of \emph{resonant} $\boldsymbol{k}$-points within the BZ
and, accordingly, we also explore the landscape of light-matter coupling
strengths, $\left|C_{\boldsymbol{k}\sigma}^{\varphi}\right|^{2}$
{[}compare to Figs.~\ref{fig:1_system} (c-f){]}. At the largest
frequencies, in Fig.~\ref{fig:3_N_Aw} (c), the region of \emph{resonant}
$\boldsymbol{k}$-points is so small {[}see Figs.~\ref{fig:1_system}
(c-d){]} that the reentrant behavior of $S^{\varphi}\left(\infty\right)$
on increasing $\tau_{\mathrm{pu}}$ is very difficult to resolve on
the 2D map, while it is clearly visible for all other high and intermediate
values of the frequency. For smaller and smaller values of the frequency,
we observe an interesting new effect: the \emph{number} of \emph{resonant}
$\boldsymbol{k}$-points with significative values of the coupling
strength becomes more and more comparable between the two spin orientations
{[}compare to Figs.~\ref{fig:1_system} (c-f){]} up to an overtaking
that reflects in an inversion of the sign of $S^{\varphi}\left(\infty\right)$
for $\omega_{\mathrm{pu}}\lesssim10\frac{t_{1}}{\hbar}$ {[}the reddish
area in Fig.~\ref{fig:3_N_Aw} (c){]}. On increasing $\tau_{\mathrm{pu}}$
in this region of pump pulse frequencies, $\omega_{pu}$, given the
strong spin dependence of the Rabi frequency, $\omega_{R,\boldsymbol{k}\sigma}^{\varphi}$,
we scan values of $\mathcal{N}_{\boldsymbol{k}\sigma}^{\varphi}\left(\infty\right)$
that again tend to compensate between the spin orientations over the
whole BZ and increase the value of $S^{\varphi}\left(\infty\right)$
up to reestablishing its positivity.

\begin{figure}[t]
\centering{\includegraphics[width=8cm]{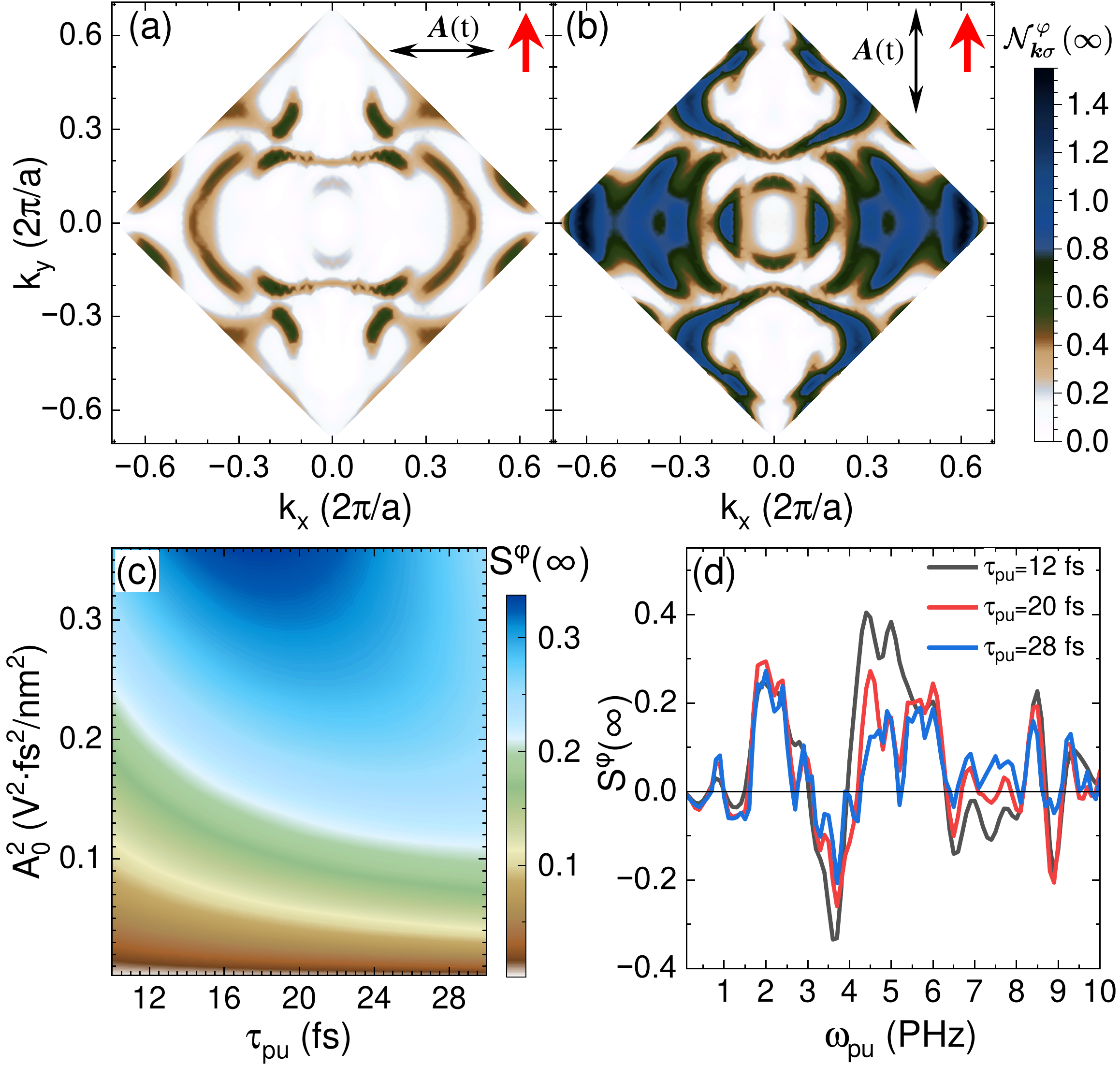}}\caption{Post-pump 
electron photo-excited populations, $\mathcal{N}_{\boldsymbol{k}\sigma}^{\varphi}\left(\infty\right)$,
for the bilayer $\mathrm{RuO}_{2}$ for (a) $\varphi=0{^{\circ}}$
and (b) $\varphi=90{^{\circ}}$. The used pump pulse parameters are
$A_{0}=2$ V~fs/nm, $\omega_{\mathrm{pu}}=2$ PHz, and $\tau_{\mathrm{pu}}=12$
fs. The temperature is fixed at about $290$ K (i.e., $25$ meV).
(c) Post-pump electron photo-excited magnetization per unit cell,
$S^{\varphi}\left(\infty\right)$, for $\varphi=90{^{\circ}}$ as
a function of (c) the FWHM, $\tau_{\mathrm{pu}}$, and the square
of the pump pulse amplitude, $A_{0}^{2}$, with $\omega_{\mathrm{pu}}=2$
PHz and (d) the pump pulse frequency, $\omega_{\mathrm{pu}}$, and
for $\tau_{\mathrm{pu}}=12,$ 20 and 28 fs with $A_{0}=2$ V~fs/nm.}
\label{fig:4_RuO2} 
\end{figure}

\textit{Pumped altermagnetic $\mathrm{RuO}_{2}$ bilayer}---To
demonstrate this phenomenology and the related effects also on the
\textit{ab initio} electronic structure of materials, with all their
complexities in terms of band structure and number of active degrees
of freedom, we performed a similar analysis on a protype AM material --
a $\mathrm{RuO}_{2}$ bilayer
(see the End Matter~C\ref{sec:RuO2-DFT}). 
While experiment evidences that bulk RuO$_{2}$ is not magnetic~\cite{hiraishi2024,liu2024absence,kiefer2024crystal,wenzel2024fermi}, altermagnetism can be stabilized in thin films~\cite{jeong2024altermagnetic,smolyanyuk2024}. 
In real materials,
charge excitations decay because of electron-phonon scattering, spontaneous
emission, etc. However, the time-scales of such processes are usually
of the order of hundreds of femtoseconds \cite{Zurch_17,PhysRevB.97.205202,marcinkevivcius2021electron,inzani2023field},
while the ultrafast pumping that we consider here occurs over a time-scale
of tens of femtoseconds. Accordingly, the post-pump photo-excitations
are computed well before such decay mechanisms become relevant. In
Figs.~\ref{fig:4_RuO2} (a) and (b), the post-pump electron photo-excited
populations, $\mathcal{N}_{\boldsymbol{k}\sigma}^{\varphi}\left(\infty\right)$,
for $\varphi=0{^{\circ}}$ and $\varphi=90{^{\circ}}$, respectively,
clearly show that through this protocol --varying the pump pulse
polarization--, it is possible to excite different spin
polarizations selectively both in specific regions of the BZ and overall in the
system. This is further confirmed by looking at the post-pump electron
photo-excited magnetization per unit cell, $S^{\varphi}\left(\infty\right)$,
as a function of the pump pulse characteristics. In Fig.~\ref{fig:4_RuO2}
(c), despite the much larger complexity of the system under analysis,
we recover the same qualitative behavior we have found for the $d$-wave
AM model {[}compare to Fig.~\ref{fig:3_N_Aw} (b){]} confirming the
robustness of the phenomenology.

In Fig.~\ref{fig:4_RuO2} (d), we see that $S^{\varphi}\left(\infty\right)$
can be easily tuned in absolute value and even in sign, on varying
both the pump pulse frequency, $\omega_{\mathrm{pu}}$, and the FWHM,
$\tau_{\mathrm{pu}}$, of the pump pulse. The complexity of \emph{real}
materials calls for different regions of the BZ for each spin orientation
to become active at the same time, in terms of being single or multi-photon
\emph{resonant} and more or less coupled to the applied EM pump pulse.
It is worth noting that, in \emph{real} materials, we can have multi-photon
resonances, and not necessarily just single-photon ones, and the former
can become even dominant with respect to the latter \citep{inzani2022field}.

The complexity of \emph{real} materials does not hinder the phenomenology,
which we proved to be very robust even in this case. Actually, such
complexity allows for greater tunability and switchability, the
possibility of designing and engineering altermagnetic materials for
specific needs, and paves the way for an effective ultrafast optical
control of AMs, with all possible future potential technological
applications.

\textit{Conclusions and Perspectives}---The phenomenology and the
effects discovered and demonstrated can be used (i) to identify altermagnetic
materials via pump-probe Kerr and Faraday spectroscopies and/or spin-
and time-resolved ARPES, (ii) to probe the spin polarization of the
band structure of AMs in energy and momentum space, (iii) to generate
photo-excited holes and electrons with a specific spin orientation
that could be used to obtain bias-induced spin-polarized currents,
and (iv) to devise an optical switch of the net spin polarization
of photo-excited holes and electrons by tuning the pump pulse characteristics.

\textit{Acknowledgments}---AE and AA
acknowledge financial support from PNRR MUR project PE0000023-NQSTI-TOPQIN.
JvdB and OJ acknowledge financial support from the Deutsche Forschungsgemeinschaft
(DFG, German Research Foundation) through the Sonderforschungsbereich
SFB 1143, grant No. YE 232/2-1, and under Germany's Excellence Strategy
through the W\"urzburg-Dresden Cluster of Excellence on Complexity and
Topology in Quantum Matter -- \emph{ct.qmat} (EXC 2147, project-ids
390858490 and 392019).

\bibliographystyle{apsrev4-2}

%

\appendix

\vspace{2em}

\begin{center}
    {\bf End Matter}
\end{center}
\subsection{A. The $d$-wave AM model\label{sec:AM-model}}

In the $d$-wave AM model \cite{antonenko2024}, the Hamiltonian is
parameterized by the Kondo interaction strength, $J$, the staggered
magnetization vector, $\boldsymbol{N}$, a nearest-neighbor hopping,
$t_{1}$, and anisotropic next-nearest-neighbor hoppings, $t_{2a}$
and $t_{2b}$, over two sublattices A and B. Then, its matrix elements
read as \cite{antonenko2024},
\begin{align}
T_{\boldsymbol{k}\nu\nu^{\prime}\sigma} & =-4t_{1}\cos\frac{k_{x}}{2}\cos\frac{k_{y}}{2}\tau_{\nu\nu^{\prime}}^{x}\nonumber \\
 & -2t_{2}\left(\cos k_{x}+\cos k_{y}\right)\tau_{\nu\nu^{\prime}}^{0}\nonumber \\
 & -2t_{d}\left(\cos k_{x}-\cos k_{y}\right)\tau_{\nu\nu^{\prime}}^{z}+JN_{z}\sigma\tau_{\nu\nu^{\prime}}^{z},
\end{align}
where $\nu,\nu^{\prime}\in\left\{ A,B\right\} $, $\tau^{0},\tau^{x},\tau^{y}$
and $\tau^{z}$ are the Pauli matrices, and $\sigma=+$ ($-$) corresponds
to the spin up (down). We set $N_{z}=\boldsymbol{N}\cdot\hat{z}=1.25\frac{4t_{d}}{J}$,
which is the staggered magnetization in the $z$ direction. Moreover,
we set $t_{2}=\frac{t_{2a}+t_{2b}}{2}=0.5t_{1}$, and $t_{d}=\frac{t_{2a}-t_{2b}}{2}=2t_{1}$.
In this model, $\boldsymbol{D}_{\boldsymbol{k}\nu\nu^{\prime}\sigma}=0$.

\subsection{B. Dynamical Projective Operatorial Approach (DPOA)\label{sec:End-material-DPOA}}

In principle, for a general time-dependent Hamiltonian $\mathcal{H}\left(t\right)$,
it is always possible to find some sets of operators, $\mathcal{C}_{\alpha}$,
that display a closed hierarchy of equations of motion: $\mathrm{i}\hbar\partial_{t}\mathcal{C}_{\alpha}\left(t\right)=\left[\mathcal{C}_{\alpha}\left(t\right),\mathcal{H}\left(t\right)\right]=\Xi_{\alpha}\left(t\right)\cdot\mathcal{C}_{\alpha}\left(t\right)$,
where $\cdot$ is the matrix product in the space of the operators
belonging to a specific set $\alpha$, while $\Xi_{\alpha}\left(t\right)$
is known as the time-dependent \emph{energy matrix} \citep{avella2011composite,eskandari2024time}.
The DPOA \citep{inzani2022field,eskandari2024time,eskandari2024generalized}
exploits this occurrence to efficiently and effectively study out-of-equilibrium
systems by projecting the time-dependent operators on their equilibrium
counterparts ($\mathcal{C}_{\alpha}\left(t\right)=P_{\alpha}\left(t\right)\cdot\mathcal{C}_{\alpha}$)
and moving the solution of the operatorial dynamics to the solution
of the equations of motion of the dynamical projection matrices $P_{\alpha}\left(t\right)$:
$\mathrm{i}\hbar\partial_{t}P_{\alpha}\left(t\right)=\Xi_{\alpha}\left(t\right)P_{\alpha}\left(t\right)$.
For a quadratic Hamiltonian of a solid-state lattice system, the operators
$\mathcal{C}$ reduce to the electronic annihilation operators.

Usually, it is convenient to work in a basis that diagonalizes the
equilibrium Hamiltonian: $T_{\boldsymbol{k}\nu\nu^{\prime}\sigma}$
in Eq.~\ref{eq:H-A-DG}. The transformation to such a basis, which
we denote by the index $n$, is performed by a unitary matrix, $\Omega_{\boldsymbol{k}\nu n\sigma}$.
The electronic annihilation operator in this basis reads as $c_{\boldsymbol{k}n\sigma}\left(t\right)$.
The corresponding eigenvalues are denoted by $\varepsilon_{\boldsymbol{k}n\sigma}$,
and are the bands of the system. One fundamental ingredient in the
DPOA framework is the use of the so-called Peierls expansion, which
is a Taylor expansion for computing very efficiently $T_{\boldsymbol{k}+\frac{e}{\hbar}\boldsymbol{A}\left(t\right)\nu\nu^{\prime}\sigma}$
and $\boldsymbol{D}_{\boldsymbol{k}+\frac{e}{\hbar}\boldsymbol{A}\left(t\right)\nu\nu^{\prime}\sigma}$
in Eq.~\ref{eq:H-A-DG} \citep{eskandari2024time}: $\chi_{\boldsymbol{k}+\frac{\mathrm{e}}{\hbar}\boldsymbol{A}\left(t\right)}\left(t\right)=\sum_{m=0}^{\infty}\frac{1}{m!}\left(\frac{\mathrm{e}}{\hbar}A\left(t\right)\right)^{m}\left(\partial_{k_{A}}\right)^{m}\chi_{\boldsymbol{k}}$,
where $k_{A}=\boldsymbol{k}\cdot\hat{\boldsymbol{A}}$ and $\hat{\boldsymbol{A}}$
is the polarization of $\boldsymbol{A}\left(t\right)=A\left(t\right)\hat{\boldsymbol{A}}$.

\subsection{C. Further details on the RuO$_{2}$ bilayer\label{sec:RuO2-DFT}}

While there is a mounting experimental evidence that bulk RuO$_{2}$ is not magnetic~\cite{hiraishi2024,liu2024absence,kiefer2024crystal,wenzel2024fermi}, altermagnetism can be stabilized in thin films~\cite{jeong2024altermagnetic,smolyanyuk2024}. Here, we consider a free-standing bilayer of RuO$_{2}$, which we use a prototype system with an effective $d$-wave AM model that mimics the complexity of \emph{real}
materials. To construct an altermagnetic bilayer, we start from the
bulk crystal structure and make a slab with two formula units
oriented along the $c$-axis of the bulk system. Opposite-spin Ru
ions are related by fourfold roto-inversion symmetry, and the system
can be represented as a $\sqrt{2}\times\sqrt{2}\times{}1$ supercell described within the space group $Cmm2$ (35) with $a=6.35$~\AA. For this structure, we perform density-functional
calculations in GGA+$U$ \cite{Perdew1997,PhysRevB.48.16929}, as
implemented in the Wien2k code~\cite{blaha2001wien2k,10.1063/1.5143061}.
We use the full localized limit for the double counting correction
and parameters $U=1.52$\,eV and $J=0.4$\,eV. For the BZ integrations,
we use a $k$-point mesh having $2000$\,$k$-points in the irreducible
BZ, along with a tetrahedron method. The calculation of the post-pump
excitations exploits a Wannier model based on Ru $4d$ and O $2p$ orbitals,
which was obtained with the Wannier90 code~\cite{Pizzi_2020}. The
model successfully reproduces the DFT band structure in the energy
range {[}-7\,eV, 3\,eV{]} around the Fermi energy.

\end{document}